\documentclass[submission,copyright,creativecommons]{eptcs}
\providecommand{\event}{PxTP 2019} 
\usepackage{breakurl}              
\usepackage{underscore}            

\usepackage[utf8]{inputenc}
\usepackage{array}
\usepackage{booktabs}

\usepackage{amssymb}
\usepackage{amsmath}
\usepackage{amsthm}
\usepackage{environ}
\usepackage{mathtools}
\usepackage[dvipsnames]{xcolor}
\usepackage{tikz}
\usepackage{siunitx}
\usepackage{listings}
\usepackage[frozencache]{minted}
\usepackage{booktabs}
\usepackage{longtable}
\usepackage{cite}
\usepackage{todonotes}
\usepackage{scalerel}
\usetikzlibrary{svg.path}

\usepackage[english]{babel}
\definecolor{orcidlogocol}{HTML}{A6CE39}
\tikzset{
  orcidlogo/.pic={
    \fill[orcidlogocol] svg{M256,128c0,70.7-57.3,128-128,128C57.3,256,0,198.7,0,128C0,57.3,57.3,0,128,0C198.7,0,256,57.3,256,128z};
    \fill[white] svg{M86.3,186.2H70.9V79.1h15.4v48.4V186.2z}
                 svg{M108.9,79.1h41.6c39.6,0,57,28.3,57,53.6c0,27.5-21.5,53.6-56.8,53.6h-41.8V79.1z M124.3,172.4h24.5c34.9,0,42.9-26.5,42.9-39.7c0-21.5-13.7-39.7-43.7-39.7h-23.7V172.4z}
                 svg{M88.7,56.8c0,5.5-4.5,10.1-10.1,10.1c-5.6,0-10.1-4.6-10.1-10.1c0-5.6,4.5-10.1,10.1-10.1C84.2,46.7,88.7,51.3,88.7,56.8z};
  }
}

\hypersetup{
   pdftitle={Reconstructing veriT Proofs in Isabelle/HOL},
   pdfauthor={Mathias Fleury and Hans-Jörg Schurr},
}
\newcommand\orcid[1]{\,\href{https://orcid.org/#1}{\mbox{\scalerel*{
\begin{tikzpicture}[yscale=-1,transform shape]
\pic{orcidlogo};
\end{tikzpicture}
}{|}}}}

\usepackage[final,tracking=true,kerning=true,spacing=true,stretch=10,shrink=10]{microtype}

\microtypecontext{spacing=nonfrench}

\newcommand{\proofRule}[1]{\microtypesetup{tracking=false}\textsc{#1}\microtypesetup{tracking=true}}
\newcommand{\isaTac}[1]{\texttt{#1}}
\newcommand\IfThenElse[3]{\ensuremath{\mathit{if}~#1~\mathit{then}~#2~\mathit{else}~#3}}

\newcommand{\proofsep}{\;\vartriangleright\;}
\newcommand{\proofsepI}{\;\blacktriangleright\;}

\NewEnviron{prooflist}{%
  \begin{spreadlines}{0.5pt}%
  \begin{align*}%
    \BODY
  \end{align*}%
\end{spreadlines}%
}

\definecolor{SmtBlue}{HTML}{00007F}
\definecolor{SmtGreen}{HTML}{3B7F31}
\newcommand{\grNT}[1]{\textcolor{SmtGreen}{\langle\texttt{#1}\rangle}}
\newcommand{\grT}[1]{\textcolor{SmtBlue}{\texttt{#1}}}
\newcommand{\grRule}{\quad\Coloneqq\quad}
\newcommand{\grOr}{\quad|\quad}

\usetikzlibrary{tikzmark,positioning}
\usetikzlibrary{shapes,arrows,fit, scopes}
\usetikzlibrary{arrows.meta}
\tikzstyle{every picture}+=[remember picture]

\tikzset{
     std/.style={draw, thin, fill=blue!20},
     system/.style={draw, thin, fill=red!20, rounded corners},
     code/.style={red, inner sep=0.15cm, text=black, draw, thick, rounded corners, fill=red!20}
}

\theoremstyle{definition}
\newtheorem{exmp}{Example}

\newcolumntype{L}[1]{>{\raggedright\let\newline\\\arraybackslash\hspace{0pt}}m{#1}}
\newcolumntype{C}[1]{>{\centering\let\newline\\\arraybackslash\hspace{0pt}}m{#1}}
\newcolumntype{R}[1]{>{\raggedleft\let\newline\\\arraybackslash\hspace{0pt}}m{#1}}

\title{Reconstructing veriT Proofs in Isabelle/HOL}
\author{Mathias Fleury\orcid{0000-0002-1705-3083}
\institute{Max-Planck-Institut für Informatik,\\ Saarland Informatics Campus, Saarbr\"ucken, Germany}
\email{mathias.fleury@mpi-inf.mpg.de}
\institute{Graduate School of Computer Science,\\ Saarland Informatics Campus, Saarbr\"ucken, Germany}
\email{s8mafleu@stud.uni-saarland.de}
\and
Hans-Jörg Schurr\orcid{0000-0002-0829-5056}
\institute{University of Lorraine, CNRS, Inria, and\\ LORIA, Nancy, France}
\email{hans-jorg.schurr@inria.fr}
}
\def\titlerunning{Reconstructing veriT Proofs in Isabelle/HOL}
\def\authorrunning{M. Fleury and H.-J. Schurr}
\begin{document}
\maketitle

\begin{abstract}
  Automated theorem provers are now commonly used within interactive
  theorem provers to discharge an increasingly large number of
  proof obligations.  To maintain the trustworthiness of a proof,
  the automatically found proof must be verified inside the proof
  assistant. We present here a reconstruction procedure in the proof assistant
  Isabelle\slash HOL for proofs generated by
  the satisfiability modulo theories solver veriT
  which is part of the \isaTac{smt} tactic.
  We describe in
  detail the architecture of our improved reconstruction method and
  the challenges we faced in designing it.  Our experiments show that
  the veriT-powered \isaTac{smt} tactic is regularly suggested by Sledgehammer
  as the fastest method to automatically solve proof goals.
\end{abstract}

\section{Introduction}
\label{sec:introduction}

\noindent
Proof assistants are used in verification, formal mathematics, and
other areas to provide trustworthy and machine-checkable formal proofs of
theorems. Proof automation allows the user to focus on the core of their
argument by reducing the burden of manual proof. A successful approach
to automation is to invoke an external automatic theorem prover (ATP),
such as a satisfiability modulo theories (SMT) solver~\cite{barrett-2009}
and to reconstruct any generated proofs using the proof assistant's
inference kernel. The usefulness of this approach depends on the encoding
of the proof goal into the language of the ATP, the capabilities of the
ATP, the quality of the generated proof output, and the reconstruction
routine itself.

In the proof assistant Isabelle/HOL this approach is implemented in
the \isaTac{smt} tactic~\cite{bohme-2010}.\footnote{Technically, \isaTac{smt}
is a proof method, but the difference (whether it requires an Isabelle context)
does not matter here.} This tactic encodes the
proof goal into the SMT-LIB language~\cite{SMTLIB} and calls the SMT
solver Z3~\cite{Z3}. If Z3 is successful in finding a proof of the
input problem, the generated proof is reconstructed inside Isabelle.
The proof format, and hence the reconstruction process, is specific to Z3.
If the reconstruction is successful, the initial proof goal holds
in the Isabelle\slash HOL logic. The reconstruction, however, might fail due to
errors (either due to a weakness in the reconstruction or due to errors to the solver) or timing out.

We have previously developed~\cite{barbosa-2019,aitp19} a prototype to reconstruct
proofs generated by the SMT solver veriT~\cite{verit}.
In this paper we have extended these works with proper support of
term sharing, tested it on a much larger scale,
and present the reconstruction method in more detail.
Furthermore, we reworked the syntax of the proof output to adhere stronger
to the SMT-LIB standard.
Given the variety of capabilities between ATPs,
a greater diversity in supported systems increases the number of
proof goals which can be solved by automated tools. Moreover, the fine-grained
proofs produced by veriT might allow for a higher success rate
in reconstruction. Lastly, the reconstruction efforts provide valuable
insights for the design of proof formats.

Similar to the proofs generated by Z3, veriT's proofs are based on
the SMT-LIB language, but are otherwise different. Proofs are a list
of indexed steps which can reference steps appearing before them in the
list. Steps without references are tautologies and assumptions. The last step is always
the deduction of the empty clause. Furthermore, steps can be marked
as subproofs, which are used for local assumptions and to reason about
bound variables. To shorten the proof length, we use term sharing,
which is implemented using the standard SMT-LIB
name annotation mechanism.  Major differences to the proof format used by Z3
are the fine-grained steps for Skolemization and the presence of
steps for the manipulations of bound variables~\cite{barbosa-2019}.

Our reconstruction routine inside Isabelle is structured as a pipeline.
Once the proof is parsed into a datatype and the term sharing is unfolded,
the SMT-LIB terms are translated into Isabelle terms.
At this point the proof can be replayed step-by-step. Special care has
to be taken to handle Skolem terms, subproofs, and the unfolding of the
encoding into the first-order logic of the SMT solver.

We validate our reconstruction approach in two ways. First, we
replace the calls of the \isaTac{smt} tactic that are currently using Z3 by veriT.
Second, we use Sledgehammer to validate the utility of veriT as a backend
solver for the \isaTac{smt} tactic. Sledgehammer uses external ATPs
to find a collection of theorems from the background theory which
are sufficient to prove the goal. It then tests a collection
of automated tactics on this set of theorems and suggests the fastest
successful tactic to the user. On theories from the Archive of Formal
Proofs, Sledgehammer suggests the usage of the veriT-powered \isaTac{smt}
tactic on a significant number of proof steps. This suggests that the
overall checking speed can be improved by switching to the veriT-powered
\isaTac{smt} tactic at these points.

\section{The Proofs Generated by veriT}
veriT is a CDCL($T$)-based satisfiability modulo theories solver.  It uses
the SMT-LIB language as input and output language and also utilizes
the many-sorted classical first-order logic defined by this language.
If requested by the user, veriT outputs a proof if it can deduce that the
input problem is unsatisfiable. In proof production mode, veriT supports
the theory of uninterpreted functions, the theory of linear integer and
real arithmetic, and quantifiers.

We assume the reader is familiar with many-sorted classical first-order
logic. To simplify the notation we will omit the sort of terms, except
when absolutely needed. The available sorts depend on the selected
SMT-LIB theory and can also be extended by the user, but a distinguished
$\mathbf{Bool}$ sort is always available.  We use the symbols $x$, $y$, $z$
for variables, $f$, $g$, $h$ for functions, and $P$, $Q$ for predicates, i.e.,
functions with result sort $\mathbf{Bool}$. The symbols $r$, $s$, $t$, $u$
stand for terms. The symbols $\varphi, \psi$ denote formulas, i.e.,
terms of sort $\mathbf{Bool}$.  We use $\sigma$ to denote substitutions
and $t\sigma$ to denote the application of the substitution on the term $t$.
To denote the substitution which maps $x$ to $t$ we write $[t/x]$.
We use $=$ to denote syntactic equality and $\simeq$ to denote the sorted
equality predicate. Since veriT implicitly removes double negations, we
also use the notion of complementary literals very liberally: $\varphi
= \bar{\psi}$ holds if the terms obtained after removing all leading
negations from $\varphi$ and $\bar{\psi}$ are syntactically equal
and the number of leading negations is even for $\varphi$ and odd for
$\bar{\psi}$, or vice versa.

A proof generated by veriT is a list of steps.  A step consists of
an index $i\in\mathbb{N}$, a formula $\varphi$, a rule name $R$ taken
from a set of possible rules, a possibly empty set of premises $\{p_1,
\dots, p_n\}$ with $p_i\in\mathbb{N}$, a rule-dependent and possibly
empty list of arguments $[a_1, \dots, a_m]$, and a context $\Gamma$.
The arguments $a_i$ are either terms or tuples $(x_i, t_i)$ where $x_i$
is a variable and $t_i$ is a term. The interpretation of the arguments
is rule specific. The context is a possible empty
list $[c_1, \dots, c_l]$, where $c_i$ stands for either a variable
or a variable-term tuple $(x_i, t_i)$. A context denotes a substitution
as described in section~\ref{sec:coreconcepts}. Every proof ends with a step
with the empty clause as the step term and empty context. The list of
premises only references earlier steps, such that the proof forms a
directed acyclic graph.
In Appendix~\ref{sec:rules} we provide an overview of all proof rules
used by veriT.

To mimic the actual proof text generated by
veriT we will use the following notation to denote a step:
\begin{equation*}
	c_1,\,\dots,\, c_l\proofsep i.\quad \varphi\quad
	(\proofRule{rule};\: p_1,\,\dots,\, p_n;\: a_1,\,\dots,\,a_m)
      \end{equation*}
If an element of the context $c_i$ is of the form $(x_i, t_i)$, we will write
$x_i\mapsto t_i$. If an element of the arguments $a_i$ is of this form
we will write $x_i \coloneqq t_i$.
Furthermore, the proofs can utilize Hilbert's choice operator $\epsilon$. Choice
acts like a binder. The term $\epsilon x. \varphi$ stands for a value $v$,
such that $\varphi[v/x]$ is true if such a value exists. Any value is possible otherwise.

The proof format used by veriT has been discussed in prior publications:
the fundamental ideas behind the proof format have been discussed
in~\cite{besson-2011}; proposed rules for quantifier instantiation can
be found in~\cite{deharbe-2011}; and more recently, veriT gained proof
rules to express reasoning typically used for processing, such as Skolemization,
renaming of variables, and other manipulations of bound variables~\cite{barbosa-2019}.
As veriT develops, so does the format of the proofs generated by it.
There also have been efforts to improve the proof generation process.
We now give an overview of the core ideas of the proofs generated by veriT
before describing the concrete syntax of the proof output.

\subsection{Core Concepts of the Proof Format}\label{sec:coreconcepts}

\paragraph{Assumptions.}
The \proofRule{assume} rule introduces a term
as an assumption. The proof starts with a number of \proofRule{assume}
steps. Each step corresponds to an assertion after some implicit transformations
have been applied as described below. Additional assumptions can
be introduced too. In this case each assumption must be discharged with
an appropriate step. The only rule to do so is the \proofRule{subproof}
rule. From an assumption $\varphi$ and a formula $\psi$ proved by
intermediate steps from $\varphi$, the \proofRule{subproof} step deduces
$\neg \varphi \lor \psi$ and discharges $\varphi$.

\paragraph{Tautologous rules and simple deduction.}
Most rules emitted by veriT introduce tautologies. One example is
the \proofRule{and\_pos} rule: $\neg (\varphi_1 \land \varphi_2 \land
\dots \land \varphi_n) \lor \varphi_i$.  Other rules operate on only
one premise. Those rules are primarily used to simplify Boolean
connectives during preprocessing. For example, the \proofRule{implies}
rule removes an implication: From $\varphi_1 \implies\varphi_2$
it deduces $\neg \varphi_1 \lor \varphi_2$.

\paragraph{Resolution.}
The proofs produced by veriT use a generalized propositional
resolution rule with the rule name \proofRule{resolution} or
\proofRule{th\_resolution}. Both names denote the same rule. The
difference only serves to distinguish if the rule was introduced by
the SAT solver or by a theory solver. The resolution step is purely
propositional; there is currently no notion of a unifier.

The premises of a resolution step are clauses and the conclusion
is a clause that has been derived from the premises by some binary
resolution steps.

\paragraph{Quantifier Instantiation.}
To express quantifier instantiation, the rule \proofRule{forall\_inst}
is used. It produces a formula of the form $(\neg \forall x_1 \dots
x_n. \varphi)\lor \varphi[t_1/x_1]\dots[t_n/x_n]$, where $\varphi$
is a term containing the free variables $(x_i)_{1\leq i\leq n}$, and $t_i$ are new
variable free terms with the same sort as $x_i$.

The arguments of a \proofRule{forall\_inst} step are the list $x_1
\coloneqq t_1, \dots, x_n \coloneqq t_n$. While this information
can be recovered from the term, providing this information explicitly aids
reconstruction because the implicit transformations applied to terms (see below) obscure
which terms have been used as instances.
Existential quantifiers are handled by Skolemization.

\paragraph{Skolemization and other preprocessing steps.}
veriT uses the notion of a \emph{context} to reason about bound variables.
As defined above, a context is a (possibly empty) list of variables
or variable term pairs. The context is modified like a stack: rules can
either append elements to the right of the current context or remove
elements from the right. A context $\Gamma$ corresponds to
a substitution $\sigma_{\Gamma}$. This substitution is recursively defined.
If $\Gamma$ is the empty list, then $\sigma_{\Gamma}$ is the empty substitution,
i.e., the identity function. If $\Gamma$ is of the form $\Gamma', x$ then
$\sigma_{\Gamma}(v) = \sigma_{\Gamma'}(v) \text{ if } v\neq x, \text{
otherwise } \sigma_{\Gamma}(v) = x$. Finally, if $\Gamma = \Gamma', x\mapsto \varphi$
then  $\sigma_{\Gamma', x\mapsto \varphi}
= \sigma_{\Gamma'}\circ [\varphi/x]$. Hence, the context allows one to build a
substitution with the additional possibility to overwrite prior substitutions
for a variable.

Contexts are processed step by step: If one step extends the context this
new context is used in all subsequent steps in the step list until the
context is modified again. Only a limited number of rules can be applied
when the context is non-empty. All of those rules have equalities as
premises and conclusion. A step with term $\varphi_1 \simeq \varphi_2$ and
context $\Gamma$ expresses the judgment that $\varphi_1\sigma_{\Gamma}
= \varphi_2$.

One typical example for a rule with context is the \proofRule{sko\_ex}
rule, which is used to express Skolemization of an existentially
quantified variable. It is a applied to a premise $n$ with a context
that maps a variable $x$ to the appropriate Skolem term and produces
a step $m$ ($m > n$) where the veriable is quantified.

\begin{prooflist}
  \Gamma, x\mapsto (\epsilon x.\varphi) \proofsep& n.&\varphi &\simeq \psi &(\proofRule{\dots})\\
  \Gamma \proofsep& m.&                 (\exists x.\varphi) &\simeq \psi &(\proofRule{sko\_ex};\: n)
\end{prooflist}

\begin{exmp}
To illustrate how such a rule is applied, consider the following example
taken from~\cite{barbosa-2019}. Here the term $\neg p(\epsilon x.\neg
p(x))$ is Skolemized. The \proofRule{refl} rule expresses
a simple tautology on the equality (reflexivity in this case), \proofRule{cong}
is functional congruence, and \proofRule{sko\_forall} works like
\proofRule{sko\_ex}, except that the choice term is $\epsilon x.\neg\varphi$.
\begin{prooflist}
  x\mapsto (\epsilon x.\neg p(x)) \proofsep& 1.&                 x  &\simeq        \epsilon x.\neg p(x)  &(\proofRule{refl})\\
  x\mapsto (\epsilon x.\neg p(x)) \proofsep& 2.&               p(x) &\simeq      p(\epsilon x.\neg p(x)) &(\proofRule{cong};\: 1)\\
                                  \proofsep& 3.&(    \forall x.p(x))&\simeq      p(\epsilon x.\neg p(x)) &(\proofRule{sko\_forall};\: 2)\\
                                  \proofsep& 4.&(\neg\forall x.p(x))&\simeq \neg p(\epsilon x.\neg p(x)) &(\proofRule{cong};\: 3)
 \end{prooflist}
\end{exmp}

\paragraph{Linear arithmetic.}

Proofs for linear arithmetic use a number of straightforward rules,
such as \proofRule{la\_totality}: $t_1 \leq t_2 \lor t_2 \leq t_1$
and the main rule \proofRule{la\_generic}.
The conclusion of an \proofRule{la\_generic} step is a tautology of the
form $(\neg \varphi_1)\lor (\neg \varphi_2)\lor\dots\lor(\neg \varphi_n)$ where the $\varphi_i$
are linear (in)equalities. Checking the validity of this formula amounts
to checking the unsatisfiability of the system of linear equations $\varphi_1,
\varphi_2, \dots, \varphi_n$.
While Isabelle provides tactics to decide the validity of a set of linear
equations, the non-trivial complexity of this task was a challenge for
the proof reconstruction (see Section~\ref{sec:arithmetic}).

\begin{exmp}
The following example is the proof generated by veriT for the unsatisfiability
of $(x+y<1) \lor (3<x)$, $x\simeq 2$, and $0\simeq y$.
\begin{prooflist}
\proofsep& 1.& &(3 < x) \lor (x + y < 1)           &(\proofRule{assume})\\
\proofsep& 2.& &x\simeq 2                          &(\proofRule{assume})\\
\proofsep& 3.& &0\simeq y                          &(\proofRule{assume})\\
\proofsep& 4.& &\neg (3<x) \lor \neg (x\simeq 2)   &(\proofRule{la\_generic})\\
\proofsep& 5.& &\neg (3<x)                         &(\proofRule{resolution};\: 2, 4)\\
\proofsep& 6.& &x + y < 1                          &(\proofRule{resolution};\: 1, 5)\\
\proofsep& 7.& &\neg (x + y < 1) \lor \neg (x\simeq 2) \lor \neg (0 \simeq y) &(\proofRule{la\_generic})\\
\proofsep& 8.& &\bot                               &(\proofRule{resolution};\: 7, 6, 2, 3)
\end{prooflist}
\end{exmp}

\paragraph{Implicit transformations.}
In addition to the explicit steps, veriT performs some transformations
on proof terms implicitly without creating steps. To ensure compatibility with
future versions of veriT, proof reconstruction must assume that those
transformations are applied between any two steps. Furthermore, veriT can not
introduce additional types of implicit transformations.

\begin{itemize}
    \item Removal of double negation: Formulas of the form $\neg (\neg \varphi)$
      are silently simplified to $\varphi$.
    \item Removal of repeated literals: If the step formula is of the form
          $\varphi_1\lor \varphi_2\lor\dots\lor\varphi_n$ with $\varphi_i
          = \varphi_j$ for some $i\neq j$, then $\varphi_j$ is removed.
          This is repeated until no more terms can be removed.
      \pagebreak[2]
    \item Simplification of tautological formulas:  If the step formula is of the form
          $\varphi_1\lor \varphi_2\lor\dots\lor\varphi_n$ with $\varphi_i
          = \bar{\varphi_j}$ for some $i\neq j$, then the formula is replaced by $\top$.
    \item Reorienting equalities: veriT applies the symmetry of equality implicitly.
\end{itemize}

\subsection{Concrete Syntax}
The concrete text representation of the proofs generated by veriT is based
on the SMT-LIB standard. Figure~\ref{fig:proof_ex} shows an exemplary proof
as printed by veriT lightly edited for readability.

We also reworked the proof syntax. Our goal is to follow the SMT-LIB standard
when possible. While those modifications do not aid reconstruction inside
Isabelle/HOL, they will simplify further development of the proof output.
Previously, veriT produced nested steps. This was changed to a flat list
of commands. The arguments of the commands are now given as annotations
instead of a flat list.
Since the changes are syntactical, the old format is still
supported by veriT and can be selected using a command-line
switch\footnote{The  option \texttt{--proof-version=N}, where \texttt{N}
is either \texttt{1}, \texttt{2}, or \texttt{3}.}.

\renewcommand{\theFancyVerbLine}{\footnotesize{\arabic{FancyVerbLine}}\hspace{-0.5em}}
\begin{figure}[t]
\begin{minted}[linenos]{smtlib2.py -x}
(assume h1 (not (p a)))
(assume h2 (forall ((z1 U)) (forall ((z2 U)) (p z2))))
...
(anchor :step t9 :args ((:= z2 vr4)))
(step t9.t1 (cl (= z2 vr4)) :rule refl)
(step t9.t2 (cl (= (p z2) (p vr4))) :rule cong :premises (t9.t1))
(step t9 (cl (= (forall ((z2 U)) (p z2)) (forall ((vr4 U)) (p vr4))))
          :rule bind)
...
(step t14 (cl (forall ((vr5 U)) (p vr5)))
          :rule th_resolution :premises (t11 t12 t13))
(step t15 (cl (or (not (forall ((vr5 U)) (p vr5))) (p a)))
          :rule forall_inst :args ((:= vr5 a)))
(step t16 (cl (not (forall ((vr5 U)) (p vr5))) (p a)) :rule or :premises (t15))
(step t17 (cl) :rule resolution :premises (t16 h1 t14))
\end{minted}
\caption{Example proof output. Assumptions are introduced (line 1--2); a subproof renames bound variables (line 4--8); the proof finishes with instantiaton and resolution steps (line 10--15)}\label{fig:proof_ex}
\end{figure}

Figure~\ref{fig:grammar} shows the grammar of the proof text generated
by veriT. It is based on the SMT-LIB grammar, as defined in the SMT-LIB
standard version 2.6 Appendix~B\footnote{Available online at: \url{http://smtlib.cs.uiowa.edu/language.shtml}}.
The nonterminals $\grNT{symbol}$, $\grNT{function\_def}$, $\grNT{sorted\_var}$,
and $\grNT{term}$ are as defined in the standard. The $\grNT{proof_term}$
is the recursive $\grNT{term}$ nonterminal redefined with the additional
production for the $\grT{choice}$ binder.

\begin{figure}[t]
  {\centering
  {
    \setlength{\belowdisplayskip}{0pt}
    \setlength{\belowdisplayshortskip}{0pt}
    \begin{align*}
      \grNT{proof}          \grRule &\grNT{proof\_command}^{*}\\
      \grNT{proof\_command} \grRule &\grT{(assume}\; \grNT{symbol}\; \grNT{proof\_term}\; \grT{)}\\
      \grOr  &\grT{(step}\; \grNT{symbol}\; \grNT{clause} \;
               \grT{:rule}\; \grNT{symbol}\;
               \grNT{step\_annotation}\; \grT{)}\\
      \grOr  &\grT{(anchor}\; \grT{:step}\; \grNT{symbol}\; \grT{)}\\
      \grOr  &\grT{(anchor}\; \grT{:step}\; \grNT{symbol}\;
               \grT{:args}\; \grNT{proof\_args}\; \grT{)}\\
      \grOr  &\grT{(define-fun}\; \grNT{function\_def}\; \grT{)}\\
      \grNT{clause}         \grRule &\grT{(cl}\; \grNT{proof\_term}^{*}\; \grT{)}\\
      \grNT{step\_annotation}\grRule&
                                      \grT{:premises (}\; \grNT{symbol}^{+}\; \grT{)}\;\\
      \grOr  &\grT{:args}\; \grNT{proof_args}\\
      \grOr  &\grT{:premises (}\; \grNT{symbol}^{+}\; \grT{)}\;
               \grT{:args}\; \grNT{proof_args}\\
      \grNT{proof\_args}    \grRule &\grT{(}\; \grNT{proof\_arg}^{+}\; \grT{)}\\
      \grNT{proof\_arg}     \grRule &\grNT{symbol} \grOr
                                      \grT{(}\; \grNT{symbol}\; \grNT{proof\_term}\; \grT{)}\\
      \grNT{proof\_term}    \grRule &\grNT{term}\text{ extended with }
                                      \grT{(choice (}\; \grNT{sorted\_var}^{+}\;\grT{)}\; \grNT{proof\_term}\; \grT{)}
    \end{align*}
  }\par}
\caption{The proof grammar}\label{fig:grammar}
\end{figure}

Input problems in the SMT-LIB standard contain a list of \emph{commands}
that modify the internal state of the solver. In agreement
with this approach veriT's proofs are also formed by a list of commands.
The $\grT{assume}$ command introduces a new assumption. While this
command could also be expressed using the $\grT{step}$ command with
a special rule, the special semantic of an assumption justifies the
presence of a dedicated command: assumptions are neither tautological nor
derived from premises.  The $\grT{step}$ command, on the other hand,
introduces a derived or tautological term.  Both commands $\grT{assume}$ and $\grT{step}$ require an
index as the first argument to later refer back to it. This index is an arbitrary SMT-LIB symbol. The
only restriction is that it must be unique for each $\grT{assume}$ and
$\grT{step}$ command.  The second argument is the term introduced by the
command.
For a $\grT{step}$, this term is always a clause.
To express disjunctions in SMT-LIB the $\grT{or}$ operator is used.
Unfortunately, this operator needs at least two arguments and cannot
represent unary or empty clauses. To circumvent this we introduce a new
$\grT{cl}$ operator.  It corresponds the standard $\grT{or}$ function
extended to one argument, where it is equal to the identity, and zero
arguments, where it is equal to $\grT{false}$.
The $\grT{:premises}$ annotation denotes the premises and is skipped
if they are none. If the rule carries arguments, the
$\grT{:args}$ annotation is used to denote them.

The $\grT{anchor}$ and $\grT{define-fun}$ commands are used for
subproofs and sharing, respectively. The $\grT{define-fun}$ command
corresponds exactly to the $\grT{define-fun}$ command of the
SMT-LIB language.

\subsection{Subproofs}
As the name suggests, the \proofRule{subproof} rule
expresses subproofs. This is possible because its application is
restricted: the assumption used as premise
for the \proofRule{subproof} step must be the assumption introduced
last. Hence, the \proofRule{assume}, \proofRule{subproof} pairs are
nested.  The context is manipulated in the same way: if a step
pops $c_1,\dots, c_n$ from a context $\Gamma$, there is a earlier step which
pushes precisely $c_1,\dots, c_n$ onto the context. Since
contexts can only be manipulated by push and pop, context manipulations
are also nested.

Because of this nesting, veriT uses the concept of subproofs.  A subproof
is started right before an \proofRule{assume} command or a command which
pushes onto the context. We call this point the \emph{anchor}.   The
subproof ends with the matching \proofRule{subproof} command or
command which pops from the context, respectively. The $\grT{:step}$
annotation of the anchor command is used to indicate the $\grT{step}$
command which will end the subproof. The term of this $\grT{step}$
command is the conclusion of the subproof.  If the subproof uses a
context, the $\grT{:args}$ annotation of the $\grT{anchor}$ command
indicates the arguments added to the context for this subproof.
In the example proof (Figure~\ref{fig:proof_ex}) a subproof starts on line four.
It ends on line seven with the $\proofRule{bind}$ steps which finished the
proof for the renaming of the bound variable \mintinline{smtlib2.py -x}{z2}
to \mintinline{smtlib2.py -x}{vr4}.

A further restriction applies: only the conclusion of a subproof can be used
as a premise outside of the subproof. Hence, a proof checking tool can
remove the steps of the subproof from memory after checking it.

\subsection{Sharing and Skolem Terms}
The proof output generated by veriT is generally large. One reason
for this is that veriT can store terms internally much more efficiently.
By utilizing a perfect sharing data structure, every term is stored in
memory precisely once. When printing the proof this compact storage
is unfolded.

The user of veriT can optionally activate sharing\footnote{By using
the command-line option \texttt{--proof-with-sharing}.} to print common
subterms only once.  This is realized using the standard naming mechanism
of SMT-LIB. In the language of SMT-LIB it is possible to annotate every
term $t$ with a name $n$ by writing $\grT{(! }t\grT{ :named }n\grT{ )}$
where $n$ is a symbol. After a term is annotated with a name, the name can
be used in place of the term. This is a purely syntactical replacement.

To limit the number of names introduced we use a simple
approach: before printing the proof we iterate over all terms of the proof
and recursively descend into the terms. We mark every unmarked subterm we
visit. If we visit a marked term, this term gets a name. If a term already
has a name, we do not descend further into this term. By doing so, we ensure that
only terms that appear as child of two different parent terms get a name.
Thanks to the perfect sharing representation testing if a term is marked takes
constant time and the overall traversal takes linear time in the proof size.

To simplify reconstruction
veriT can optionally\footnote{By using the command-line option
\texttt{--proof-define-skolems}.} define Skolem constants as functions. If activated,
this option adds a list of $\grT{define-fun}$ command to define shorthand
0-ary functions for the $\grT{(choice }\dots\grT{)}$ terms needed. Without
this option, no $\grT{define-fun}$ commands are issued and the constants are inlined.

\section{Proof Reconstruction in Isabelle/HOL}
\label{sec:proof-rec}

\begin{figure}[t]
\centering
\scalebox{0.8}{
\begin{tikzpicture}[node distance=4cm, auto,>=latex', thick,scale=0.8]
	\begin{scope}
	    \path[->] node[system] (verit) {veriT proof};
	    \path[->] node[std, right=4cm of verit] (ushp) {Unshared structured proof}
	                  (verit) edge node[align=left] {
            \begin{minipage}{3cm}\raggedright
              Parse raw proof$^\star$ and unfold sharing
            \end{minipage}} (ushp);
	    \path[->] node[std, below=1.2cm of ushp] (ppp) {Preprocessed proof}
	                  (ushp) edge node[align=left] {Convert to Isabelle/HOL terms$^\star$\\ and add dependencies} (ppp);
	    \path[->] node[system, left=5cm of ppp] (rply) {Proof} (ppp) edge node[above,name=ed3] {Replay} (rply);
	\end{scope}

	\begin{scope}[yshift=-4.5cm, xshift=1.5cm]
          \path[->] node[system] (proof) {List of steps};
          \node[std, right=4cm of proof] (hrp)  {Replayed Proof};
          \node[system, below=1cm of hrp] (bot) {$\bot$};
          \path[->] (proof) edge[below] node {
            \begin{minipage}{3.5cm}
              Unfold FO encoding\\Replay other steps$^{\dag}$
            \end{minipage}
          } (hrp);
           \path[->] (hrp) edge [right] node (ed1){Discharge Skolems} (bot);
          \node (box) [dotted, rounded corners=1mm, fit = (ed1)(bot)(proof) (hrp), draw] {};
          \draw[thick, dotted] (box.north west) -- ([yshift=-1mm]ed3.south west);
          \draw[thick, dotted] (box.north east) -- ([yshift=-1mm]ed3.south east);
	\end{scope}

    \node at (19.3,0) {
      \begin{minipage}{0.3\linewidth}\raggedright
        $^\star$shared with the reconstruction for Z3~\cite{bohme-2010}.
      \end{minipage}
    };
    \node at (19.3,-1.7) {
      \begin{minipage}{0.3\linewidth}\raggedright
        $^{\dag}$only resolution is shared with Z3~\cite{bohme-2010}.
        Most methods changed from~\cite{barbosa-2019}.
      \end{minipage}
    };
\end{tikzpicture}}
\caption{The reconstruction pipeline}\label{fig:isa-pipe}
\end{figure}
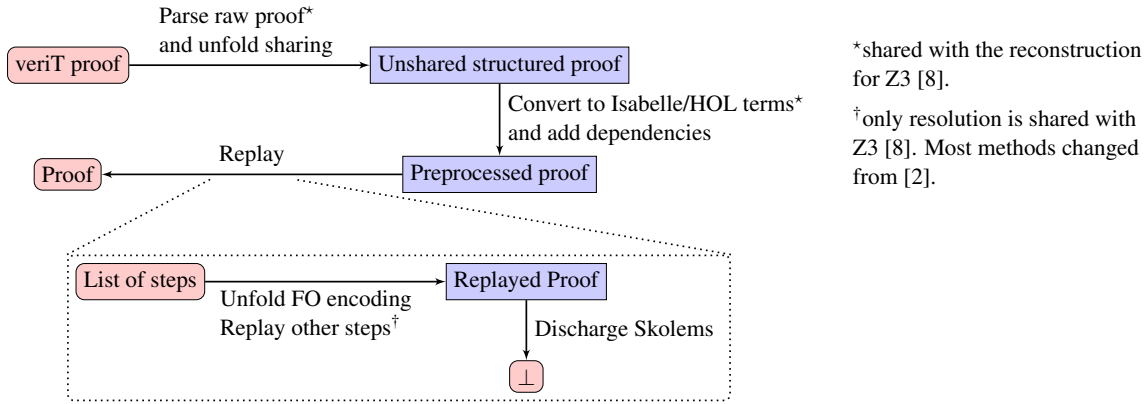

Proof reconstruction is done in Isabelle in two steps presented in
Figure~\ref{fig:isa-pipe}: first, the proof is parsed and the terms are transformed
into Isabelle terms (Section~\ref{sec:pars-prepr}). Then we can reconstruct
the proof itself by reconstructing the steps one-by-one. Some of the
steps require some care (Section~\ref{sec:reconstructing-proof}).

\subsection{Parsing and Preprocessing}
\label{sec:pars-prepr}

Parsing the proof is simple thanks to the infrastructure developed to
reconstruct Z3 proofs for the \isaTac{smt} tactic. This infrastructure is able to parse a
generalized version of the SMT-LIB syntax, including the proofs generated by veriT.
It produces a raw version of the proof. We only have to extract the
structure (indices, steps, \ldots) from the raw proof. During parsing of the raw proof
we also unfold the sharing, because Isabelle does not offer any sharing functionality.

The first transformation is a change of the disjunction representation. In the proof output,
veriT represents
the outermost disjunction as a multiset by using the $\grT{cl}$ operator. In Isabelle, we
replace this multiset
by a disjunction.
veriT explicitly applies the rule \proofRule{or} to convert disjunctions to multisets. In Isabelle,
these steps are simply the identity.

After that, we convert the SMT-LIB terms to Isabelle terms. This reuses
again some of the infrastructure developed for Z3. An important difference with Z3 is the
declaration of variables. When converting to Isabelle terms, types are inferred.
However, some expressions like $x \mapsto y$ of the context cannot be typed
without extracting the types from the conclusion.

Finally, we preprocess the proofs to ease the reconstruction further:
\begin{itemize}
\item We add the implicit dependency between the last step of
  each subproof and its conclusion. In the example of Figure~\ref{fig:proof_ex},
  it is the dependency from
  \mintinline{smtlib2.py -x}{t9} to \mintinline{smtlib2.py -x}{t9.t2}.
  Spelling it out explicitly makes the reconstruction more regular.
\item We add missing dependencies to the definitions of Skolem terms: veriT
  applies definitions implicitly, but we have to unfold the definitions
  explicitly to reconstruct Skolemization steps in Isabelle.
\end{itemize}

The Isabelle semantics of the proof steps are slightly different than
the semantics in veriT. First, the context is seen as a list of equalities
instead of a list of mappings. Second, the conclusion uses the
equality symbol instead of $\simeq$ and a substitution.  If the proof step is
$ y\mapsto z, x\mapsto s \proofsep n.\;\varphi\simeq \psi$, Isabelle sees it as:
$ y = z, x = s\proofsepI n.\;\varphi =  \psi$, where
the context are assumptions.
The advantage of this different semantics is that it requires fewer transformations on the input
term, as it avoids adding lambda abstractions, and makes the Isabelle
tactics easier to use for reconstructions.

The difference in the semantics is small and rarely important. They
only differ for variables that can syntactically appear on
the left and on the right-hand side with different semantics.
For example, consider $y\mapsto z, x\mapsto y \proofsep n.\;P\,x\,y\simeq P\,y\,z$.
This is a tautology, because $P\,x\,y$ is $P\,y\,z$ after substitution (remember
that the substitution only applies on the left-hand side). However, the
naive conversion to the Isabelle version yields
$y = z, x = y \proofsepI n.\;P\,x\,y\simeq P\,y\,z$, which is
a different term, namely $P\,z\,z = P\,z\,z$

To avoid the difference in semantics, we rename terms when they have already been
bound: we rename the occurrences on the right-hand side of $\simeq$ of
$y$ by the new fresh name $\mathit{xy}$. The step
$ y\mapsto z, x\mapsto y \proofsep n.\;\varphi\simeq \psi$ becomes
$ y = z, x = \mathit{xy}\proofsepI n.\;\varphi = \psi[\mathit{xy}/x]$.

\subsection{Reconstructing Parsed Proofs}
\label{sec:reconstructing-proof}

After parsing, we reconstruct the proof steps in Isabelle. 
Overall, the proof reconstruction works by replaying each
step and unifying the assumptions with the premises. At the end, we get a
proof of $\bot$.

For most steps,
the rule can be spelled out as an Isabelle theorem and the only issues are implicit
steps (Section~\ref{sec:application-theorems}). Unlike the
reconstruction of Z3 proofs, we reconstruct subproofs as they are printed by veriT
(Section~\ref{sec:subproofs}). Finally,
some rules require special care or are tricky to reconstruct:
Skolemization steps, if done naively, can produce terms that are too
large to be handled efficiently
(Section~\ref{sec:skolemizations}); Isabelle's arithmetic procedure is
incomplete and not very efficient when reconstructing arithmetic
steps (Section~\ref{sec:arithmetic}); for
efficiency, some rules are reconstructed heuristically
(Section~\ref{sec:other-rules}).

\subsubsection{Application of Theorems}
\label{sec:application-theorems}

Most rules that can be applied are either tautologies or applications
of theorems that can be easily expressed: The rule \proofRule{true}
is the tautology used to prove that the theorem $\top$ holds. Similarly,
the transitivity rule \proofRule{eq\_transitive} transforms the assumptions
$(t_j \simeq t_{j+1})_{j < n}$ into $t_0 \simeq t_n$.

In practice, there are two main difficulties: double negations can be
simplified and equalities can be reoriented. The reorientation is implementation
dependent, which prohibits us from relying on the order
as given by the input problem. The reordering and simplification have consequences that
either make reconstruction harder or require additional annotations in
the proof output:
\begin{itemize}
    \item Duplicate literals are implicitly removed. This is
      rarely an issue in practice, but we have seen this happening in
      some test cases like the \proofRule{ite2} rule. This rule introduces
      the tautology $(\IfThenElse{\neg \varphi}{\psi_1\\}{\psi_2})\lor \varphi \lor \psi_2$, but
      if $\varphi = \psi_2$ the it produces the simplified clause
      $(\IfThenElse{\neg \varphi}{\psi_1}{\psi_2})\lor \varphi$.
    \item The rules can be applied up to additional negations. For
      example, the \proofRule{ite2} rule can
      be applied to get
      $(\IfThenElse{\varphi}{\psi_1}{\psi_2})\lor (\neg \varphi) \lor \psi_2$.
\end{itemize}

\noindent
The individual steps are reconstructed by:
\begin{itemize}
\item taking into account the additional information provided
  in the proof output. This can require some preprocessing on the
  formula: in veriT instantiation (rule \proofRule{forall_inst}) can be done to
  quantifiers that do not appear
  at the outermost level, but inside the formula. Preprocessing is
  used to transform $\forall x. (P \implies \neg(\exists y. Q\,y))$ into
  $\forall x\, y.\, (P \implies Q\,y)$. This is easier to reconstruct,
  because all quantifiers to instantiate are now at the outmost level and
  forall quantifiers.
\item applying the theorem  or  finding the instantiations and then using
  \isaTac{simp} to reorder the equalities and prove that the terms are
  equal. For the \proofRule{ite2} on
  $(\IfThenElse{\neg \varphi}{\psi_1}{\psi_2})\lor \varphi' \lor \psi_2'$,
  we identify the terms, $\varphi$, $\psi_1$, and $\psi_2$, and generate
  the tautology $(\IfThenElse{\neg \varphi}{\psi_1}{\psi_2})\lor \varphi \lor \psi_2$,
  that can be used by \isaTac{simp} to discharge the goal by showing $\varphi = \varphi'$ and
  $\psi_2=\psi_2'$.
  The search space is very large and the search can be very time consuming during the reconstruction.
\item providing various version of the lemmas to accommodate negations:
  for Isabelle, the theorem $(\IfThenElse{\neg \varphi}{\psi_1}{\psi_2})\lor \varphi \lor \psi_2$
  cannot be applied to prove
  $(\IfThenElse{\varphi}{\psi_1}{\psi_2}) \lor \neg \varphi \lor \psi_2$.
\end{itemize}

In practice the reordering happens mostly when producing new terms (during parsing or
instantiation). However, we do not want to rely on this specific
behavior which could change in a future version.

\subsubsection{Subproofs}
\label{sec:subproofs}

Unlike Z3, veriT has subproofs. Subproofs fall into two categories:
proofs used to justify proof steps (e.g.\ for Skolemization) and lemmas
with assumptions and fixed variables. In Isabelle, both are modeled by
the notions of contexts that encapsulate all the assumptions and fixed
variables present at a given point.

The first kind of subproofs are proofs of \emph{lemmas} that come with additional
assumptions. They are used for example for proofs like $P \implies
\bot$. $P$ is an assumption of the proof (given by an \proofRule{assume}
command) and $\bot$ is the conclusion.
In Isabelle, we start by extracting all the assumptions when entering the
proof. This creates a new context. Then, we replay the proof in the new
context. The \proofRule{assume} commands are now entailed
by the context are replayed as such. Finally, the conclusion is exported to the outer context.

Replaying \emph{subproofs} is similar to replaying lemmas in the proof: we
enter contexts with new assumptions and variables, depending on the rule. At the end of the subproof
the last step is exported back to the outer context and is used to discharge
the conclusion. For example, the subproof of a simple \proofRule{bind} step will be of the form
$\forall x\,y.\,(x = y \implies P\,x = Q\, y)$ to prove that $(\forall x.\, P \,x) = (\forall y.\, Q \,y)$

\subsubsection{Skolemizations}
\label{sec:skolemizations}

Skolemization is an important but subtle point, which slightly differs
between Isabelle and veriT. While defining the constants is easy,
the definitions themselves do not exactly match the natural ones
and reconstructing the proof can be difficult.

Technically, Skolem constants are not introduced with a definition, but
as an assumption of the form $\overline{X} = (\epsilon x.\, \dots)$. At the end
of the reconstruction, we
get the theorem
$\forall \overline{X}.\, (\overline{X} = (\epsilon x. \neg P\, x) \implies \bot)$,
from which we can trivially derive the theorem $\bot$.

Internally, veriT directly Skolemizes formulas: The term
$\forall x\,y.\; P\,x\,y$ becomes after Skolemization 
$P\, \overline{X}\,\overline{Y}$, i.e., $ P\,\left(\epsilon x. \neg (\forall y.\; P\,x\,y) \right)\,
\left(\epsilon y. \neg P\,\left(\epsilon x. \neg (\forall y.\;
    P\,x\,y) \right) \right)$, where $\overline{X}$ and $\overline{Y}$ are defined to be the two Skolem
constants.
However, in the logged proof,
$\forall x\,y.\; P\,x\,y$ becomes
  $\forall y.\;P\,\overline{X}\, y$, i.e., $\forall y.\;P\,\left(\epsilon x. \neg (\forall y.\; P\,x\,y)
  \right)\, y$, which in turns naturally becomes
  $P\,\overline{X}\,
  \left(\epsilon y. \neg P\,\overline{X}\,y \right)$.
Therefore, in Isabelle, we fold the
definition of the Skolems inside each other to get
$\overline{Y} = \left(\epsilon y. \neg P\,\overline{X}\,y \right)$ and try to prove the goal.
This might, however, fail due to the implicit steps. Hence,
if required, we unfold all definitions and prove the result.
This could explode for non-trivial terms, but we did not have issues with this during our experiments.

A major issue of the reconstruction is the size of generated terms. While
developing the reconstruction, we found a case where four variables
were Skolemized in a single step, and the generated term was so big that
Isabelle was not able to replace the third variable by the equivalent
choice: the application of the theorem
$(\forall x. \, P\,x) \iff P\,(\epsilon x. \neg P\,x)$ was too slow. 
We now aggressively
fold the Skolem constants inside the term.

\subsubsection{Arithmetic}
\label{sec:arithmetic}

To replay arithmetic steps, we use Isabelle's procedure
\isaTac{linarith}. This tactic is a decision procedure for real
numbers, but not for integers or natural numbers. Internally, it uses
the Fourier–Motzkin elimination~\cite{DBLP:books/daglib/0090562}: it derives a contradiction
via a linear combination of the equations.

veriT with proof production only supports linear arithmetic. On linear
problems, however, it is stronger than Isabelle's tactic: Isabelle does
not simplify equations. If we have the equations $5 \times x + 10 \times y \simeq 15$, it will
not be simplified to $x + 2 \times y \simeq 3$ in Isabelle. This happens neither as preprocessing,
nor during the search for the linear combination. In one case over the
Archive of Formal Proofs, this makes the following problem impossible to reconstruct:

\[
    \begin{array}{ccccccccr}
      \neg& 0& \leq& y & \land \quad
      \neg& 10\times x &<&4 + 14\times z\quad &\land\\
      &10\times x&\le&15+25\times y & \land\quad
      \neg& 10\times  x + 10\times  z &\le&30+25\times y &
    \end{array}
\]

\noindent This goal is produced as an arithmetic tautology by veriT,
but \isaTac{linarith} is not able to prove it. Before simplification,
the inequality $16 \le 10 \times  x - 25 \times  y$ is derived. After simplification, the equivalent
(but seemingly stronger) inequality $20 \le 10 \times  x - 25 \times  y$ is
derived because $x$ and $y$ are integers.
The coefficients of the second inequality are different enough to allow
\isaTac{linarith} to find a contradiction,
which it was unable to find otherwise.

We strengthened the reconstruction by implementing a
simplification procedure that divides each equation by its greatest
common divisor. It could be activated more globally, but currently
conflicts with two other simplification procedures: one of them sorts
terms, while the other does not.

\subsubsection{Other Rules}
\label{sec:other-rules}

The reconstruction of the rules is often guided by the efficiency of
the reconstruction, how often a rule is used, and concrete examples.
One of the most prominent rules is \proofRule{connective\_equiv}. It is a
simplification step and can involve simplifications of the Boolean
structure and arithmetic. At first, we reconstructed \proofRule{connective\_equiv} steps with
\isaTac{auto}, a tactic that simplifies the terms and performs some logical
reasoning. However, this turned out to be too inefficient on large
terms. Moreover, often only the Boolean structure is modified and
not the terms or the order of equalities. Therefore, we now
first abstract over the non-Boolean terms and check only the
modifications on the Boolean structure by \isaTac{fast}. Only if
this fails is \isaTac{auto} tried. If that also fails, \isaTac{metis} is
tried as a fallback tactic. We do not attempt to select the right tactic,
but simply try them in this order.

\section{Experimental Results}

We experiment on the Isabelle reconstruction in two
ways. The first one is to replace all the \isaTac{smt} calls
that are in the Isabelle distribution
and are currently powered by Z3 by the version of \isaTac{smt} with veriT. These
\isaTac{smt} calls have
been selected by the developer of the library who provided the theory, because Z3 is able to find a proof and
the reconstruction is fast. While this experiment
provides insight into the performance of the veriT-powered \isaTac{smt}
tactic relative to the Z3-powered variant, it does not tell us if
the veriT-powered one is a useful and supplementary addition to the
family of automated tactics provided by Isabelle.
Towards that end, we try to
generate new veriT-powered \isaTac{smt} calls by using Sledgehammer~\cite{Blanchette2016-isar}, an
Isabelle tool able to find proofs.

\subsection{Replacing the \isaTac{smt} calls}
\label{sec:smt-calls}

\begin{table}[t]
  \centering
  \begin{tabular}{lr}
    SMT calls                   & Number of occurrences\\ \midrule
    Successful reconstruction   & 447\\
    Failed reconstruction       &   4\\
    veriT timeouts              &  47\\
    veriT unknown               &   4
  \end{tabular}
  \caption{Result of using veriT instead of Z3 in existing \isaTac{smt} calls}
  \label{tab:smt-calls}
\end{table}

There are already many \isaTac{smt} calls in the theories included
in the Isabelle distribution and the Archive
of Formal Proofs. The latter did not allow \isaTac{smt} calls until a few years
ago. We replaced the Z3 as a backend for the \isaTac{smt} calls,
by veriT. The results are summarized in
Table~\ref{tab:smt-calls}. Testing revealed:
\begin{itemize}
\item that veriT is not able to find all the proofs that Z3 is able to find. 
  On the one hand, this does not corroborate the findings from the
  SMT competition where veriT performs better than Z3 on some categories.
  On the other hand, the problems have been specifically
  selected to be solvable by Z3. We did not include the problems
  specifically relying on Z3 extensions (e.g. the division operator) or
  features not supported by veriT (bit vectors).
\item a bug in the proof generation. veriT does not
  correctly print some substeps: a term is replaced by an equivalent
  term, but this replacement is not logged. We are currently
  fixing this bug in veriT. In Isabelle, this leads to an error in the
  reconstruction and we do not attempt to reconstruct the following steps.
\item the problem in the reconstruction of
  arithmetic steps described in Section~\ref{sec:arithmetic}.
  Only one of those benchmarks could not be reconstructed without the simplification procedure.
\end{itemize}

The results are promising: we are able to reconstruct nearly all of
the proofs that veriT is able to find. We cannot replace Z3 by
veriT in the Isabelle distribution, but this was not the aim of this
experiment.

\subsection{Generating calls with Sledgehammer}
\label{sec:sledgehammer-tests}

\emph{Hammers}, like Sledgehammer, select facts from the background theory,
translate them to the input language of the provers, and then attempt
to use the generated proof if a proof is found within a
given timeout. The proof can be used in different ways. One approach is to
gather the facts required to find a proof with
one of the builtin tactics of the proof assistant. Another approach is to replay the proof
within the core by an \isaTac{smt}-like method or to translate it
into the user-facing language of the assistant.
Sledgehammer supports all of these approaches.

By default, Sledgehammer uses the following strategy: first, it tries several
Isabelle tactics, including detailed proof reconstruction
by the Z3-powered \isaTac{smt} tactic, with a timeout of \SI{1}{\second}.
If this is successful, it returns
the tactic that was the fastest to prove the goal in Isabelle. This tactic can
be inserted in the theory. If none
of them is fast enough to find a proof, reconstruction of a
proof in the user-facing language is attempted. We have
changed Sledgehammer to additionally test the veriT-powered
\isaTac{smt} tactic. Sledgehammer tries to find a proof and to minimize
the involved facts. Even when Z3 finds the proof,
reconstruction with veriT can be faster.

We tested this approach on two formalizations:
an ordered resolution
prover~\cite{Ordered_Resolution_Prover-AFP,DBLP:conf/cade/SchlichtkrullBT18}
and the SSA
language~\cite{DBLP:conf/cc/BuchwaldLU16,Formal_SSA-AFP}.
We tested all theories included in the formalization.
We selected the first development because
we knew that Sledgehammer was
useful during the development.  We selected the second formalization because it
is a very different theme. Due to time constraints, we did
not test more theories.

The results are given in Table~\ref{tab:sledgehammer-stats}. They show
that veriT-powered \isaTac{smt} calls happen in practice and can improve the
speed of the overall proof processing. We do not know why
veriT performs much worse on the formal SSA theory, but we believe
that some rules that we do not reconstruct efficiently enough
(possibly the \proofRule{qnt_simplify} rule that simplifies quantifiers) appear more
often in this theory.
The row `Oracle' denotes calls to solvers that found a proof that could not be reconstructed.
Many of these failed calls are proofs found by the SMT solver CVC4
that can either not be found or not be reconstructed by veriT and Z3-powered \isaTac{smt}.

\begin{table}[tb]
  \centering
  \begin{tabular}{p{4cm}rr}
    Theory                           &Ordered Resolution Prover&Formal SSA\\
    \midrule
    Found proofs                     & 5019                    &5961\\
    Z3-powered \isaTac{smt} proofs   &   90                    & 109\\
    veriT-powered \isaTac{smt} proofs&   25                    &   4\\
    Oracle       &    9                    &  63
  \end{tabular}

  \caption{Proofs found by Sledgehammer on two Isabelle formalizations}
  \label{tab:sledgehammer-stats}
\end{table}

\section{Related Work}
\label{sec:related-work}

Reconstruction of proofs generated by external theorem provers has
been implemented in
various systems including CVC in HOL Light~\cite{McLaughlinBG06}, Z3
in HOL4 and Isabelle/HOL~\cite{bohme-2010}, and SMTCoq reconstructs veriT~\cite{armand-et-al-2011} and
CVC4~\cite{EkiciKKMRT16} proofs in Coq. None
of the other solvers produce detailed proofs or information on Skolemization.
For veriT proofs, SMTCoq currently supports a different version of the proof output
(version~1) that has different rules and an older version of veriT (the version is from 2016),
which does not record detailed information for Skolemization and
has worse performance.

The reconstruction of Z3 proofs in HOL4 and Isabelle/HOL is one of the
most advanced and well tested. It has been used to check proofs
generated on problems from the SMT competition. Sadly, the code
to read the SMT-LIB input problems was
never included in the standard Isabelle distribution and is now lost.
Proof reconstruction has been heavily tested and
succeeds in more than 90\% of the cases according to Sledgehammer
benchmark~\cite[Section~9]{Blanchette2016-isar}, and is very efficient.

The SMT solver CVC4 follows a different philosophy from veriT
and Z3: it produces proofs in a logical framework with side
conditions~\cite{Stump-2013}.  The output can contain
programs to check certain rules.  The CVC4 proof format is  quite
flexible but currently CVC4 does not produce proofs
for quantifiers.

\section{Conclusion and Future Work}

We presented the syntax and semantics of the proofs generated by veriT and the
reconstruction of those proofs in Isabelle. During the development, the format
was extended to ease reconstruction
by printing more information like the instantiations. We hope to integrate our
code in the next Isabelle release.

Overall, having more details in the proofs helps to make the
reconstruction more robust, because each step is simpler to check. For
example, veriT detailed information on Skolemization, makes it easier
to replay than the one from Z3: the reconstruction can call the
ordered resolution prover \isaTac{metis}. For now, the
implicit simplifications prevents us from reconstructing proof more
efficiently than Z3.

Another challenge is to translate the proof to the more readable Isar
format. It is useful for two main reasons. First, it gives the
Isabelle user more information on how the proof works and potentially
what kind of lemmas would be interesting to create. Second, if the
reconstruction fails, it allows the user to fix the
failing part. Generating readable proofs can be done automatically by
Sledgehammer for most solvers, but this does not work for veriT proofs. One reason is that,
Sledgehammer does not support subproofs and inlining the assumptions
each time is not very readable. Another reason is that the Skolems constants implicitly
introduce a context where these constants are defined. This introduces
an implicit dependency order between the definitions and every step where the defined constant
appears. We could unfold the definitions to use the choice version instead, but that
would harm the readability of the proof. Finally, 
the proofs generated by veriT often follow the scheme ``$\varphi$ holds;
$\varphi \leftrightarrow \psi$ also holds; hence
$\psi$ holds'', whereas ``$\varphi$ hence $\psi$'' is easier to
understand.

There are various useful pieces of information that are found by the solver but are not
presented to the user. For example, in the case of linear arithmetic a
contradiction is derived by finding a linear
combination of the equations, but the coefficients are not printed. Therefore,
Isabelle must find these same coefficients
again. The reconstruction would be faster if they were in the proof
output.

\paragraph{Acknowledgments.}

We thank Alex Brick, Daniel El Ouraoui, and Pascal Fontaine for suggesting many textual improvements.
The work has received funding from the European Research Council (ERC)
under the European Union's Horizon 2020 research and innovation program
(grant agreement No. 713999, Matryoshka).
Previous experiments were carried out
using the Grid'5000 testbed (\url{https://www.grid5000.fr/}), supported by
a scientific interest group hosted by Inria and including CNRS, RENATER,
and several universities as well as other organizations.
\bibliographystyle{eptcs}
\bibliography{bib}

\appendix

\section{List of Proof Rules}
\label{sec:rules}

\renewcommand{\arraystretch}{1.2}
  \begin{longtable}{@{\hspace{1pt}}L{10cm}L{5.6cm}@{\hspace{1pt}}}
 \toprule
 Rule & Description\\ \midrule\midrule

 \proofRule{true}, \proofRule{false}, \proofRule{and_pos}, \proofRule{and_neq},
 \proofRule{or_pos}, \proofRule{or_neg},
 \proofRule{implies_pos}, \proofRule{implies_neg1}, \proofRule{implies_neg2},
 \proofRule{equiv_pos1}, \proofRule{equiv_pos2}, \proofRule{equiv_neg1},
 \proofRule{equiv_neg2}, \proofRule{ite_pos1}, \proofRule{ite_pos2},
 \proofRule{ite_neg1}, \proofRule{ite_neg2}, \proofRule{eq_reflexive},
 \proofRule{refl},\proofRule{trans},\proofRule{cong}, \proofRule{not_or},
 \proofRule{implies}, \proofRule{not_implies1},
 \proofRule{not_implies2}, \proofRule{equiv1},\proofRule{equiv2},
 \proofRule{not_equiv1}, \proofRule{not_equiv2}, \proofRule{ite1},
 \proofRule{ite2}, \proofRule{not_ite1}, \proofRule{not_ite2},
 \proofRule{ite_intro}                                              & Simple rules without premises \\\hline
 \proofRule{or}, \proofRule{bind}, \proofRule{eq_transitive}        & Simple rules with premises \\\hline

 \proofRule{eq_congruent}, \proofRule{eq_congruent_pred}            & Congruence. Reconstruction can be problematic due to the FO encoding. \\\hline

 \proofRule{la_rw_eq}, \proofRule{la_generic}, \proofRule{lia_generic},
 \proofRule{la_disequality}, \proofRule{la_totality},
 \proofRule{la_tautology}                                           & Linear arithmetics \\\hline
  
 \proofRule{forall_inst}                                            & Variable instantiation \\\hline
    \proofRule{th_resolution}, \proofRule{resolution}                  & Resolution reconstructed with a simple SAT solver in Isabelle
    \\\hline
 \proofRule{connective_equiv}                                       & Arithmetics and logic simplification \\\hline
 \proofRule{tmp_ac_simp}                                            & Simplification modulo associativity and commutativity \\\hline
 \proofRule{subproof}                                               & Implication from assumption\\\hline
 \proofRule{sko_ex}, \proofRule{sko_forall}                         & Skolemization \\\hline

 \proofRule{qnt_simplify}, \proofRule{qnt_join},
 \proofRule{qnt_rm_unused}, \proofRule{tmp_bfun_elim}               & Quantifier simplification\\\hline
 \proofRule{let}, \proofRule{xor1}, \proofRule{xor2},
 \proofRule{not_xor1}, \proofRule{not_xor2}, \proofRule{xor_pos1},
 \proofRule{xor_pos2}, \proofRule{xor_neg1}, \proofRule{xor_neg2},
 \proofRule{distinct_elim}                                          & Unused: Isabelle does not generates XOR or lets\\\hline
 \proofRule{nla_generic}, \proofRule{tmp_skolemize}                 & Unused: experimental features\\
 \bottomrule
 \end{longtable}
\end{document}